\documentclass[twocolumn,pre,nofootinbib]{revtex4}

\usepackage{amsmath,amssymb,bm,amsthm}
\usepackage{epsfig}

\let\emptyset\varnothing
\newcommand{\y}{y}

\begin{document}
\title{Spontaneous genetic clustering in populations of competing organisms}
\author{Tim Rogers$^1$, Alan J. McKane$^1$ and Axel G. Rossberg$^2$}\address{$^1$Theoretical Physics Division, School of Physics \& Astronomy, The University of Manchester, M13 9PL, UK\\ $^2$ Centre for Environment, Fisheries and Aquaculture Science (Cefas), Pakefield Road, Lowestoft, Suffolk NR33 0HT, UK}
\begin{abstract}
We introduce and analyse an individual-based evolutionary model, in which a population of genetically diverse organisms compete with each other for limited resources. Through theoretical analysis and stochastic simulations, we show that the model exhibits a pattern-forming instability which is highly amplified by the effects of demographic noise, leading to the spontaneous formation of genotypic clusters. This mechanism supports the thesis that stochasticity has a central role in the formation and coherence of species.
\end{abstract}
\maketitle
\section{Introduction}
The development of a quantitative theory of speciation is of fundamental biological importance, however, the complex relationships between the various mechanisms at work make this enterprise fraught with difficulties. The analysis of simple mathematical models of evolutionary dynamics can provide invaluable insight, particularly as a tool to distinguish necessary and sufficient conditions for the formation of new species. In recent years there has been considerable interest in the possibility of sympatric speciation driven by competition for resources. The mathematical formulation of this problem dates back to MacArthur and Levins \cite{MacArthur1967}, although similar models have been proposed by many others \cite{Sasaki1997,Dieckmann1999,Fuentes2003,Hernandez-Garcia2004,Scheffer2006,Pigolotti2007}. The robustness of this mechanism of speciation has been called into question, however, as species may or may not form depending on the precise details of how the effects of competition are modelled \cite{
Polechova2005,Pigolotti2010,Fort2010}. 

Traditionally, many mathematical models of evolutionary processes are formulated at a macroscopic level, describing the dynamics of entire populations and neglecting the effects of intrinsic demographic noise. In a recent study \cite{Rogers2012} we analysed an individual-based (i.e., micro-level) stochastic model of competition between individual organisms. The model reduces to the usual population-level equations in the limit of infinite population size but, crucially, for finite populations we found that speciation is dramatically enhanced by the effects of demographic noise. This observation serves firstly to show that competition-driven speciation is in fact far more robust an effect than is suggested by deterministic analyses. Secondly, it illustrates the need to take the individual nature of the organisms into account when modelling speciation.

The models discussed above all relate to phenotypic speciation, where phenotypes are represented by a numerical value in a one-dimensional `niche-space' \footnote{A notable exception is \cite{Maruvka2008}, where a complex network of possible organism types is considered.}. This is, of course, an over-simplification, and care should be taken in drawing conclusions on the basis of such models. In this paper, we introduce and analyse a genetic counterpart to the individual-based model studied in \cite{Rogers2012}, with the purpose of investigating speciation from a different, and complementary, viewpoint. Our model consists of a population of organisms which are characterised by their ``genomes'', which we model as binary sequences. The organisms reproduce (with mutation) and die due to competition between individuals, where competition is strongest between organisms with similar genomes.  

We are interested in studying the formation of species, which we interpret as well-separated clusters of genetically similar organisms. One immediate problem which arises is the question of how to detect such clusters from the genetic data. This is precisely the problem faced by biologists seeking to classify organisms using genetic sequencing and many methods exist \cite{Durbin1998,Deonier2005}. For our analysis, we choose to study the distribution of genetic distance between pairs of randomly selected organisms. This statistical measure is sufficient to determine if genetic clusters have formed, as well as being amenable to theoretical analysis. Moreover, closely related measures have already been employed in experimental genetics, for example \cite{Goldenfeld2012}. 

Starting from the individual-based stochastic model, we perform a systematic expansion in population size, providing a mathematical description of the model at three levels. In the limit of infinite populations we recover a deterministic system of differential equations for the frequency of different genomes in the population. Analysis of this system reveals a pattern-forming instability which may be interpreted as describing the formation of disjoint genetic clusters, that is, the formation of species. For large but finite populations, a linear noise analysis shows that demographic stochasticity acts to enhance the clustering process, leading to quasi-clusters in an otherwise homogeneous population. Finally, a full (non-perturbative) analysis is possible in the neutral case of global competition. Depending on the scaling relationship between mutation rate and population size, we find that demographic noise can lead to the population spontaneously forming sharply delineated genetic clusters.

The paper is organised as follows. The model is defined in the next section, after which Section III deals with the mathematical reformulation of the model in terms of a Langevin equation defined in a high-dimensional space. The theoretical analysis of deterministic, weak- and strong-noise effects is presented in Section IV, along with comparisons to simulation data. In Section V we conclude with a discussion of our findings. There are two technical appendices.

\section{A genetic model of competition}
In this section we introduce an individual-based stochastic model of a competing population. The genetics of this population are modelled by using binary sequences of length $N$; in our biological analogy, a zero or a one at a given point in this sequence tells us which of two possible gene variants (alleles) is present at that locus \footnote{Alternatively, our binary sequences can be thought of as a reduced model of DNA, with two nucleotides instead of four.}. An individual organism is specified by its genome or, more conveniently, by the list of positions of the type-one alleles it possesses. For example, the 8-bit genome $(0,1,1,0,0,0,1,0)$ corresponds to the set $\{2,3,7\}$. At time $t$ in the model, there are $\mathcal{N}(t)$ living organisms, with genomes labelled by sets $I_1,\ldots,I_{\mathcal{N}(t)}$.\par
It is also necessary to define a notion of genetic similarity between organisms. We choose to measure the distance between two genomes by counting the number of entries they have in common, known as the Hamming distance \cite{Hamming1950}. This is a standard approach in quantifiying genetic distance in experimental studies, for example \cite{Goldenfeld2012}. The Hamming distance between genomes labelled with sets $I$ and $J$ is equal to the number of elements appearing in one of $I$ or $J$, but not both. We will use the notation $I\ominus J=\{n\,\,:\,\,n\in I\cup J\,\,\,\textrm{and}\,\,n\notin I\cap J\,\}$, so that the Hamming distance between $I$ and $J$ may be written $|I\ominus J|$, where $|\,\cdot\,|$ denotes the cardinality of the set.\par
Each organism reproduces asexually with the same constant rate. We choose our timescale so as to set this rate to one. The genome of the offspring is cloned from that of the parent, with the possibility of some mutation: each point in the gene sequence has a probability $\mu$ of flipping between 0 and 1. We consider all sequences among the $2^N$ possible combinations to be viable, so each reproduction event results in the addition of an organism to the population. The rate with which an organism with genome $I$ gives birth to one with genome $J$ is thus
\begin{equation}
R_{IJ}=\mu^{|I\ominus J|}(1-\mu)^{N-|I\ominus J|}\,.
\label{defR}
\end{equation}\par
Deaths in our model result from competitive interactions between the organisms. In phenotypic models of competition, it is assumed that individuals with similar phenotypes are likely to exploit their environment in similar ways, and will thus compete more with each other than with organisms whose phenotypes are very different. We apply the same convention to our genotypic model, with the assumption that the map between genotype and phenotype is sufficiently simple that we may treat competition as a function of genetic similarity. We define the strength of competition between organisms with genomes $I$ and $J$ to be a function of their Hamming distance:
\begin{equation}
G_{IJ}=g(|I\ominus J|)\,.
\label{defG}
\end{equation}
The function $g$ is chosen to be decreasing (so that competition strength declines with genotypic distance), and normalised according to
\begin{equation}
\frac{1}{2^N}\sum^{N}_{n=0} {N \choose n} g(n) = 1.
\label{normg}
\end{equation}
This particular choice of normalisation is made in order to simplify the expression for the overall carrying capacity of the system, as will be made clear later.

The death rate of organism $n$ at time $t$ is given by the total competition it experiences, multiplied by a constant $\kappa$. This parameter controls the carrying capacity: when $\kappa$ is large, competition is fierce and only a few organisms can coexist; when it is small, death rates are low and the population grows large. In fact this relationship is rather precise; it can be seen from both the simulations and theory that the total population is typically close to $1/\kappa$. \par
The birth and death rates defined above specify the dynamics of the model. Starting from an initial seed population consisting of $\mathcal{N}(0)=1/\kappa$ organisms with uniformly randomly assigned genomes, we allow the processes of reproduction and competition to shape the population. For numerical simulations this is achieved using Gillespie's algorithm \cite{Gillespie1977}.

\section{Mathematical formulation}
\subsection{Master equation}
We now embark on a theoretical analysis of the behaviour of our model of genetic competition. The first step is to formulate the model in the standard way as a Markov process described by a master equation \cite{vanKampen1992}.\par
At time $t$, we specify the state of the system by a vector $\bm{x}$ with entries indexed by the subsets of $\{1,\ldots,N\}$. The entry $x_I$ gives the (scaled by $\kappa$) number of organisms with genome $I$:
\begin{equation*}
x_I=\kappa\sum_{n=1}^{\mathcal{N}(t)}\delta_{I_n,I}\,.
\end{equation*}
Our analysis concerns the time evolution of the distribution $P(\bm{x},t)$, giving the probability of finding the system in state $\bm{x}$ at time $t$. To determine the rate of change of $P$ in time, we must consider contributions coming from the two processes which alter the system state -- birth and death.\par
The birth of an organism with genome $I$ alters the state of the system through the addition of $\kappa$ to $x_I$. The rate $B_I$ with which this event occurs is found by summing the birth rate of all existing organisms (which we have set equal to unity) multiplied by the probability of the offspring being suitably mutated to have genome $I$. That is,
\begin{equation*}
B_I =\sum_{n=1}^{\mathcal{N}(t)}R_{II_n}=\sum_{J}\,\sum_{n=1}^{\mathcal{N}(t)}R_{I J}
\delta_{I_n ,J} = \frac{1}{\kappa}\sum_{J}R_{IJ}x_J\,.
\end{equation*}
The death rate of an organism with genome $I$ is given by the sum of the competition between itself and the other organisms, multiplied by $\kappa$. Multiplying this quantity by the number of organisms with that genome (i.e. $x_I/\kappa$) gives a total death rate of
\begin{equation*}
D_I=x_I\sum_{n=1}^{\mathcal{N}(t)}G_{II_n}=\frac{1}{\kappa}\sum_{J}x_IG_{IJ}x_J\,.
\end{equation*}
Combining the effects of these two processes, we may write the master equation as \cite{vanKampen1992}
\begin{eqnarray}
\frac{d P}{d t}&=\sum_I\bigg\{\Big(\mathcal{E}_I^--1\Big)B_IP +\Big(\mathcal{E}_I^+-1\Big)D_IP\Big]\bigg\}\nonumber\\
&=\frac{1}{\kappa}\sum_{I,J}\bigg\{\Big(\mathcal{E}_I^--1\Big)\Big[R_{IJ}x_JP\Big]\nonumber\\
&\qquad\qquad\qquad+\Big(\mathcal{E}_I^+-1\Big)\Big[G_{IJ}x_Ix_JP\Big]\bigg\}\,,
\label{meqn}
\end{eqnarray}
where $\mathcal{E}_I^{\pm}$ is a step operator which alters its argument through the addition of $\pm \kappa$ to $x_I$. 
\subsection{Kramers-Moyal expansion}
We are interested in the limit of small $\kappa$, in which the effect of competition is weak and hence the population grows large. In this regime we approximate $P$ by a continuous probability distribution $\mathcal{P}$, and expand the step functions in their Taylor series:
\begin{equation}
\mathcal{E}_I^{\pm}=\sum_{i=0}^\infty (\pm \kappa)^i\frac{\partial^i}{\partial x_I^i} \,.
\end{equation}
Applying this expansion to the master equation (\ref{meqn}) and truncating at $i=2$ yields the non-linear Fokker-Planck equation \cite{Gardiner1985}
\begin{eqnarray}
\frac{\partial \mathcal{P}}{\partial t}=&-\sum_{I,J}\frac{\partial}{\partial x_I}\bigg[\Big(R_{IJ}x_J-G_{IJ}x_Ix_J\Big)\mathcal{P}\bigg]\nonumber\\
&+\frac{\kappa}{2}\sum_{I,J}\frac{\partial^2}{\partial x_I^2}\bigg[\Big(R_{IJ}x_J+G_{IJ}x_Ix_J\Big)\mathcal{P}\bigg]\,.
\end{eqnarray}
For our purposes, it will be more convenient to work with the equivalent Langevin equation (using the It\={o} formalism) \cite{Gardiner1985}:
\begin{eqnarray}
\frac{dx_I}{dt}=&\sum_J \big(R_{IJ}x_J-x_IG_{IJ}x_J\big)\nonumber \\
&+\,\left[\kappa\sum_J \big(R_{IJ}x_J+x_IG_{IJ}x_J\big)\right]^{1/2}\eta_I(t)\,,
\label{langevin}
\end{eqnarray}
where the $\eta_I(t)$ are independent Gaussian white noise variables with zero mean and unit variance, that is, $\langle \eta_I(t)\eta_J(t')\rangle=\delta(t-t')\delta_{I,J}$. Here, and hereafter, we use $\langle\cdots\rangle$ to denote averaging over the noise.  \par
It is worth pausing for a moment at this stage to discuss the precise sense in which equation (\ref{langevin}) describes the behaviour of our original microscopic stochastic model. The astute reader may be concerned by the fact that we treat the $x_I$ as continuous stochastic variables, when in reality the large number of possible genomes means that most $x_I$ will be exactly zero, with only a few taking values $\kappa$, $2\kappa$, etc. The explanation is that, although $P$ and $\mathcal{P}$ take different arguments (one discrete, the other continuous), their first and second order moments agree up to $\mathcal{O}(\kappa^{2})$. The error committed rigorously bounded by Kurtz \cite{Kurtz1978}; as we will see, this approximation is quite sufficient for our purposes.
\subsection{An orthogonal basis}
The simulations presented in later sections are taken from a model with an $N=32$ bit genome. Even with this relatively low number of loci, the system (\ref{langevin}) has some 4,294,967,296 dimensions. Care needs to be taken to arrive at analytical results which are computationally tractable. The first simplifying step we take is to change basis with the aim of diagonalising the mutation and competition matrices $R$ and $G$, defined in (\ref{defR}) and (\ref{defG}). \par
We will be making use of the discrete Fourier transformation on the space of binary sequences. To do this, we introduce the matrix $\Pi_{IJ}=(-1)^{|I\cap J|}$ and the transformation
\begin{equation}
\widetilde{\bm{f}}_I=\sum_J \Pi_{IJ}\bm{f}_J\,.
\end{equation}
Using Eq.~(\ref{invPi}) of Appendix A, the inverse transformation is 
\begin{equation}
\bm{f}_I=\frac{1}{2^N}\sum_J \Pi_{IJ}\widetilde{\bm{f}}_J\,.
\label{invf}
\end{equation}

The most useful property of the matrix $\Pi$ is that it diagonalises Hamming-distance invariant functions. Generally, if $F$ is a matrix with entries $F_{IJ}=f(|I\ominus J|)$ for some function $f$, then $\Pi F\Pi$ is diagonal. Proof of this fact is given in Appendix A. Both $R$ and $G$ matrices have this property and
so are completely characterised by the quantities
\begin{equation}
\rho_I\equiv\frac{1}{2^N}\Big[\Pi R \Pi \Big]_{II}\quad\textrm{and}\quad \gamma_I\equiv\frac{1}{4^N}\Big[\Pi G \Pi \Big]_{II}\,,
\end{equation}
for $I\subseteq\{1,\,\ldots\,,N\}$. It is shown in Appendix A that $\rho_{I}= (1-2\mu)^{|I|}$, which implies that $\rho_{\emptyset}=1$ for all $\mu$. It also follows that 
\begin{equation*}
\gamma_{I} = \frac{1}{2^N}\,\sum_{J} \Pi_{IJ} g(|J|)\,,
\end{equation*}
and so $\gamma_{\emptyset} = 2^{-N}\,\sum_{J} g(|J|)$, since the entries of $\Pi_{\emptyset,J}$ are all equal to unity. Fixing $\gamma_{\emptyset}$ specifies the normalisation of the competition kernel. A convenient choice is $\gamma_{\emptyset}=1$ which, since there are ${N \choose n}$ genomes with $|J|=n$, gives the normalisation specified in Eq.~(\ref{normg}).

The useful properties of this transform motivate a change of variables from $\bm{x}$ to $\bm{y}=\widetilde{\bm{x}}$. Carrying this out in Eq.~(\ref{langevin}) we arrive at the Langevin equation
\begin{equation}
\frac{d\y_I}{dt}=\y_I\,\rho_I-\sum_J\y_J\,\y_{I\ominus J}\,\gamma_{I\ominus J}+\sqrt{\kappa}\,\zeta_I(t)\,.
\label{langeviny}
\end{equation}
Here the $\zeta_I(t)$ are Gaussian noise variables with correlations 
\begin{eqnarray}
&\big\langle \zeta_I(t)\,\zeta_J(t')\big\rangle\Big.\nonumber\\
&\quad=\delta\big(t-t'\big)\sum_{K,L}\Pi_{IK}\big( R_{KL} x_L + x_IG_{KL}x_L \big) \Pi_{KJ}\nonumber\\
&\quad=\delta\big(t-t'\big)\bigg(\y_{I\ominus J}\rho_{I\ominus J}+\sum_{K}\y_{K}\,\y_{I\ominus J\ominus K}\gamma_{I\ominus J\ominus K}\bigg)\,.
\label{noisey}
\end{eqnarray}
Equations (\ref{langeviny}) and (\ref{noisey}) will form the starting point for our analysis of the behaviour of the system.
\section{Analysis}
\subsection{Deterministic dynamics}
We first consider the behaviour of the model in the limit of very large population sizes, with mutation strength held constant. This corresponds to taking $\kappa\to 0$, in which case the Langevin equation (\ref{langeviny}) reduces to the deterministic system
\begin{equation}
\frac{dy_I}{dt}=y_I\,\rho_I-\sum_Jy_J\,y_{I\ominus J}\,\gamma_{I\ominus J}\,.
\label{deterministic}
\end{equation}
Now, since $\rho_\emptyset=\gamma_\emptyset=1$, we find that the deterministic equation (\ref{deterministic}) has a fixed point at $y_I=\delta_{\emptyset,I}$. In the original variables, this corresponds to $x_I=2^{-N}$ for all $I$; that is, the organisms are spread homogeneously throughout the genetic space. We denote by $A$ the Jacobian matrix of (\ref{deterministic}) at this fixed point, whose entries are
\begin{equation}
A_{IJ}=\delta_{I,J}\big(\rho_I-\gamma_I-1\big)\,.
\label{jac}
\end{equation}
The homogeneous fixed point is therefore stable if and only if $\rho_I-\gamma_I<1$ for each $I$. The boundary of the stability of the homogeneous state is determined by the balance between the strength of mutation and the shape of competition kernel. To illustrate this, we consider a particular choice of kernel with a `top-hat' shape parameterised by the width $w\in[0,N]$. Let
\begin{equation*}
g(n)=\Bigg\{\begin{array}{c}1/g_w\quad\textrm{if}\quad n\leq w\\0\qquad\,\,\,\textrm{otherwise,} \end{array} 
\end{equation*}
where $g_w$ is the normalisation constant enforcing (\ref{normg}). \par
\begin{figure}
\begin{center}
\includegraphics[width=0.48\textwidth, trim=0 0 20 10, clip]{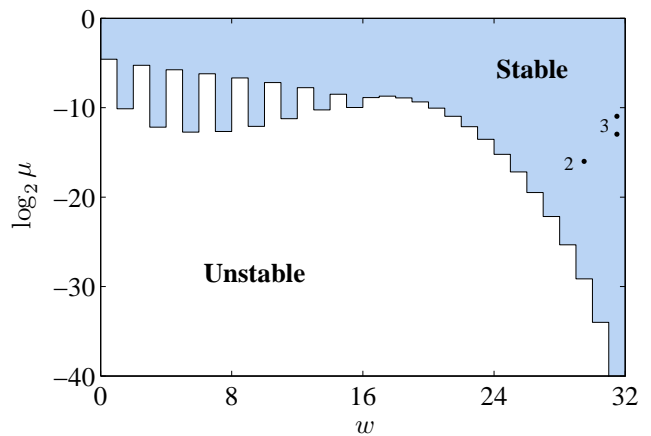}
\end{center}
\caption{Phase diagram showing the stability of the homogeneous state of the 32-bit genome model in the deterministic limit $\kappa\to 0$. The parameters $w$ and $\mu$ on the horizontal and vertical axes respectively control the width of the top-hat competition kernel and the strength of mutation. Numbered dots show the parameter values used for simulations appearing in later figures (with corresponding figure numbers). \label{phased}}
\end{figure}

Figure \ref{phased} shows the phase diagram in this case with axes for mutation strength $\mu$ and kernel width $w$. The unusual sawtooth shape of the boundary may be attributed to the geometry of sequence space, since the overlap between top-hat competition kernels (i.e. spheres) depends on their parity. The packing of spheres in sequence space is itself a difficult problem in information theory, with roots going back to the seminal work of Hamming on error-correcting codes \cite{Hamming1950}.\par
What behaviour will the model exhibit in the unstable regime? If the system is unstable in direction $y_I$ then the population density variables $x_J$ will each either be exponentially enhanced or suppressed, according to the sign of $\Pi_{IJ}$. This is a pattern-forming instability in direct analogue with those occurring in spatial systems \cite{Pigolotti2007} and on networks \cite{Maruvka2008}. Once a pattern has formed, some clusters of genomes will be very common amongst the population while others are totally absent. This process can be thought of as describing the formation of species: the population has split into several groups which are genetically isolated from each other.\par
We should point out that not all choices of kernel will result in a pattern-forming transition in the deterministic dynamics. From the earlier stability analysis, we see that if $\gamma_I>0$ for all $I$, then the homogeneous state is always stable and clusters cannot form. This result is quite restrictive, as several simple choices of kernel (for example, one of a similar form to the reproductive kernel) satisfy this condition and therefore appear not to result in clusters. The same situation is found in the traditional setting of a one-dimensional niche space, where it has been found to rule out the overlap of resource consumption as being responsible for cluster formation \cite{Roughgarden1979}. However, we will see that the effects of demographic noise are powerful enough to override this analysis. 
\subsection{Weak noise effects}
In our previous work on phenotypic competition, we found that demographic noise strongly affected the formation of clusters \cite{Rogers2012}. It is natural to ask if the same is true in the present genome-based model. \par
As a first approximation, we look for small stochastic corrections to the deterministic system. Suppose we are in the situation that the homogeneous state $y_I=\delta_{\emptyset,I}$ is stable in the deterministic dynamics (\ref{deterministic}). We linearise the Langevin equation (\ref{langeviny}) around this state, introducing the change of variables $z_I=\big(y_I-\delta_{I,\emptyset}\big)/\sqrt{\kappa}$. Keeping only the lowest order terms in $\kappa$ we arrive at the linear stochastic differential equation
\begin{equation}
\frac{d\bm{z}}{dt}=A\bm{z}+\sqrt{2}\,\bm{\xi}(t)\,,
\label{langevinz}
\end{equation}
where $A$ is the Jacobian matrix defined in (\ref{jac}) and $\bm{\xi}(t)$ is a vector of independent white noise variables with unit variance. This is an Ornstein-Uhlenbeck process,  whose general solution is known \cite{vanKampen1992}. \par
For our purposes, we are mainly interested in the behaviour of the correlations between variables. Let us define the shorthands $Y_{IJ}=\langle y_Iy_J \rangle$ and $Z_{IJ}=\langle z_Iz_J \rangle$. A standard result \cite{vanKampen1992} states that if $\bm{z}$ satisfies (\ref{langevinz}), then for $Z$ we have
\begin{equation*}
\frac{dZ}{dt}=AZ+ZA^T+2\,\mathbb{I}\,,
\end{equation*}
where $\mathbb{I}$ denotes the identity matrix of size $2^N$. Since in our case the matrix $A$ is diagonal, the dynamics of the $Z_{IJ}$ are independent of one another and can be solved easily. In particular, in the long time limit we find
\begin{equation*}
Z_{IJ}\to\,\delta_{I,J}\,\frac{1}{1+\gamma_I-\rho_I}\,,
\end{equation*}
and thus, changing back to $y$ variables, 
\begin{equation}
Y_{IJ}\to\,\delta_{I,J}\,\left(\delta_{I,\emptyset}+\frac{\kappa}{1+\gamma_I-\rho_I}\right)\,.
\label{YIJ}
\end{equation}
The $\delta_{I,\emptyset}$ part in the above equation comes from the deterministic part of the $y$ variables, and the second term from the stochastic corrections, which are of order $\sqrt{\kappa}$. \par
It is not immediately obvious from (\ref{YIJ}) what qualitative difference to the genetics of the population will result from this stochastic term. To help answer this question, we investigate the distribution of genetic (Hamming) distance between randomly selected organisms. For each $n\in\{0,\ldots,N\}$, define
\begin{eqnarray}
\Xi(n)&=\kappa^2\sum_{k,l}\delta_{|I_k\ominus I_l|,n}=\sum_{I,J} \delta_{|I\ominus J|,n} x_Ix_J\,.
\label{defXi}
\end{eqnarray}
If two organisms are selected at random from the population, $\Xi(n)$ gives the probability that their genomes differ in $n$ loci. It is straightforward to compute that at the deterministic fixed point $x_I\equiv 2^{-N}$ the shape of $\Xi$ is a symmetric binomial distribution: $\Xi(n)= 2^{-N}{N \choose n}$. \par
The calculation of the covariance of Fourier variables $y$ gives sufficient information to compute the long-time average form of $\Xi(n)$ in the presence of noise. From Eq.~(\ref{defXi}): 
\begin{eqnarray}
\big\langle\Xi(n)\big\rangle_\infty &= \sum_{I,J} \delta_{|I\ominus J|,n} \langle x_Ix_J \rangle_\infty\nonumber\\
&=2^N{N \choose n}\big\langle x_\emptyset x_{\{1,\ldots,n\}} \big\rangle_\infty\nonumber\\
&=\frac{1}{2^N}{N \choose n}\sum_{I,J}\Pi_{I,\emptyset}\Pi_{J,\{1,\ldots,n\}}\,Y_{IJ}\,,
\label{XiY}
\end{eqnarray}
where $\langle\cdots\rangle_\infty$ refers to averaging over the stationary distribution. The second equality comes from the symmetry between genomes, meaning that we may choose to study the pair $I=\emptyset$ and $J=\{1,\ldots,n\}$, which is representative of all $2^N{N \choose n}$ pairs of Hamming distance $n$.\par
In the regime of weak noise, the long-time behaviour of the correlation function is given by Eq.~(\ref{YIJ}). Using this result in Eq.~(\ref{XiY}) gives
\begin{equation}
\big\langle\Xi(n)\big\rangle_\infty = \frac{1}{2^N}{N \choose n} + \kappa 
\sum_{I} \frac{\Pi_{I,\emptyset}\Pi_{I,\{1,\ldots,n\}}}{1+\gamma_{I}-\rho_{I}},
\label{weak_xi}
\end{equation}
which clearly shows the deterministic result plus the order $\kappa$ stochastic correction. \par
\begin{figure}
\begin{center}
\includegraphics[width=0.48\textwidth, trim=0 5 10 10, clip]{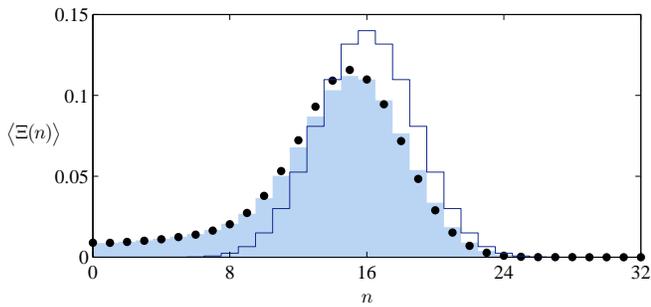}
\end{center}
\caption{Distribution of pairwise Hamming distance as measured from simulations in the weak noise regime (black circles) and predicted by the theory (blue/grey area). The binomial distribution predicted by the deterministic theory is shown for comparison (line). Parameters here are $w=30$, $\mu=2^{-16}$, $\kappa=10^{-3}$, and the simulation result was the averaged over 100 samples at taken time $t=1000$. \label{weakfig}}
\end{figure}

A typical example of weak noise affecting the distribution of Hamming distances is shown in Fig.~\ref{weakfig}. The theoretical prediction from Eq.~(\ref{weak_xi}) is compared with data gathered from simulations, averaged over 100 samples. We have chosen a top-hat competition kernel, the phase diagram for which is given in Fig.~\ref{phased}. The parameters $w=30$, $\mu=2^{-16}$, $\kappa=10^{-3}$ are well within the region of stability for the homogeneous state, meaning that the deterministic theory predicts that the distribution of pairwise Hamming distance should be binomial. As is visible in Fig.~\ref{weakfig}, there is a significant noise-induced deviation: the distribution is skewed to the left, that is, randomly selected organisms often have more genetic data in common than one would expect. Demographic noise is causing the formation of genotypic clusters.

\subsection{Strong noise effects} 
Moving beyond weak noise effects, we can make further analytic progress by considering the paradigmatic `neutral' case in which competition strength is independent of genetic distance and thus all organisms have equal fitness. This is a special case of the top-hat kernel we considered earlier, with $w=N$ and thus the homogeneous state is stable for all values of the mutation coefficient $\mu$. \par
Changing basis, $g(n)\equiv1$ gives $\gamma_I=\delta_{I,\emptyset}$, and thus the Langevin equation for the $y$ variables simplifies to 
\begin{equation}
\frac{dy_I}{dt}=y_I\,\big(\rho_I-y_{\emptyset}\big)+\sqrt{\kappa}\,\zeta_I(t)\,,
\label{oldsystem}
\end{equation}
where
\begin{equation}
\big\langle \zeta_I(t)\,\zeta_J(t')\big\rangle=\delta\big(t-t'\big)\,y_{I\ominus J}\big(\rho_{I\ominus J}+\,y_{\emptyset}\big)\,.
\label{oldnoise}
\end{equation}
Notice that the dynamics of $y_\emptyset$ are separated from those of the other variables: we have 
\begin{equation*}
\frac{dy_\emptyset}{dt}=y_\emptyset(1-y_\emptyset)+\sqrt{\kappa}\,\zeta_\emptyset(t)\,,
\end{equation*}
where
\begin{equation*}
\big\langle \zeta_\emptyset(t)\zeta_\emptyset(t')\big\rangle=\delta(t-t')\,y_\emptyset(1+y_\emptyset)\,.
\end{equation*}
This equation describes noisy logistic growth, and the long-time quasi-stationary distribution was computed in \cite{Rogers2012}. Unsurprisingly, as $\kappa\to0$, the distribution of $y_\emptyset$ approaches a delta function centred on one. \par
We can exploit this fact mathematically through the use of adiabatic elimination, setting $y_\emptyset\equiv 1$ and thus $\zeta_\emptyset(t)\equiv 0$. In Appendix B we derive general expressions for conditioned stochastic differential equations, which can be applied here to give for $I,J\neq\emptyset$
\begin{equation}
\frac{dy_I}{dt}=y_I\,\big(\rho_I-1\big)+\sqrt{\kappa}\,\zeta_I(t)\,,
\label{langevinS}
\end{equation}
where now
\begin{eqnarray}
&\big\langle \zeta_I(t)\zeta_J(t')\big\rangle=\nonumber\\
&\delta\big(t-t'\big)\left(y_{I\ominus J} \big(\rho_{I\ominus J}+1\big)-y_Iy_J\, \frac{\big(\rho_{I}+1\big)\big(\rho_{J}+1\big)}{2}\right)\,.
\label{condy}
\end{eqnarray}
Comparing this expression to (\ref{oldnoise}) we see that conditioning on the value of $\zeta_\emptyset$ results in an anti-correlation between the other noise variables which previously was not present. This will act to enhance the formation of certain patterns of clusters. \par
We are now in a position to evaluate the dynamics of the moments of the remaining degrees of freedom. From Eq.~(\ref{langevinS}), we have
\begin{equation*}
\frac{d\langle y_{I}\rangle}{dt} = (\rho_{I}-1) \langle y_{I}\rangle,
\end{equation*}
for $I \neq \emptyset$. Since $\rho_I<1$ for all $I$, each $\langle y_{I}\rangle$ undergoes exponential decay. Earlier we specified that the genomes of the initial `seed' population are randomly assigned, we thus deduce that the relation 
\begin{equation}
\langle y_{I}\rangle = \delta_{I,\emptyset}
\label{mean_of_y}
\end{equation}
holds throughout. Moving on to examine the covariance structure, we employ It\={o}'s lemma \cite{Ito1951} to obtain the following equation for $Y_{IJ}=\langle y_{I}y_{J}\rangle$ with $I,J \neq \emptyset$:
\begin{eqnarray}
\frac{dY_{IJ}}{dt} &=& \bigg(\rho_{I}+\rho_{J}-2-\frac{\kappa}{2}\,\big(\rho_{{I}}+1\big)\big(\rho_{{J}}+1\big)\bigg)Y_{IJ} \nonumber \\
&&+\, \kappa\,\big(\rho_{{I}\ominus {J}}+1\big)\langle y_{{I}\ominus {J}} \rangle\,.
\label{YIJdot}
\end{eqnarray}
Substituting for $\langle y_{{I}\ominus {J}} \rangle$ using Eq.~(\ref{mean_of_y}), we find that the only non-zero contributions to $Y_{IJ}$ arise when ${I} \ominus {J} = \emptyset$, that is, when $I=J$. Solving Eq.~(\ref{YIJdot}), we find that in the long-time limit 
\begin{equation*}
Y_{IJ}\to\delta_{{I},{J}}\,\left[\left(\frac{\rho_{I}+1}{2}\right)^2-\frac{\rho_{I}-1}{\kappa}\right]^{-1}\,.
\end{equation*}
Recalling that $\rho_I=(1-2\mu)^{|I|}$, we observe that the scale of $Y_{II}$ is determined by the relationship between competition strength $\kappa$ and mutation rate $\mu$. In the two limiting cases; we have $Y_{II}=0$ when $\kappa=0\,,\,\mu\neq0$ and $Y_{II}=1$ when $\kappa\neq0\,,\,\mu=0$. \par
\begin{figure}
\begin{center}
\includegraphics[width=0.45\textwidth, trim=0 10 25 0, clip]{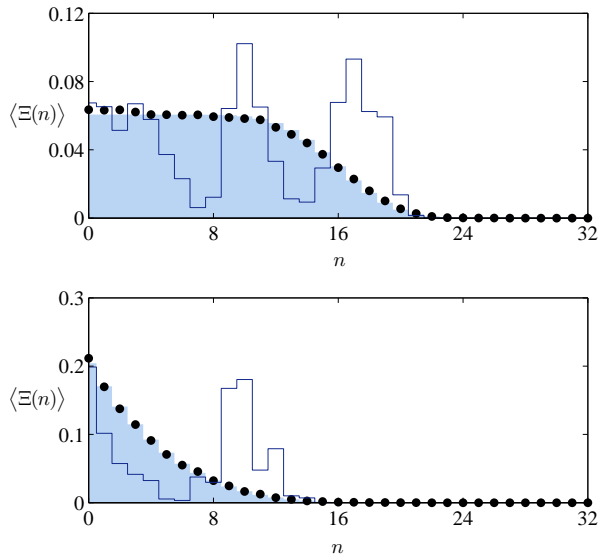}
\end{center}
\caption{Distribution of pairwise Hamming distance as measured from simulations in the strong noise regime (black circles) and predicted by the theory (blue/grey area). The parameter values are $\kappa=10^{-3}$ for both plots, $\tau=1$ in the upper plot and $\tau=4$ in the lower. The simulation result was averaged over 1000 samples taken at time $t=10000$. The unaveraged data is highly random; in both plots the dark line shows the distribution of pairwise Hamming distance measured from the first simulation in the sample. \label{strongfig}}
\end{figure}
We can explore the range between these extremes by taking the limit $\kappa\to0$ and $\mu\to0$ with $\tau \equiv \kappa/2\mu$ fixed. Biologically, this corresponds to the joint scaling in which populations are very large and mutations very rare, but the total number of mutations per generation occurring in the whole population remains approximately constant. In this case the above equation simplifies to 
\begin{equation}
Y_{IJ} \to \delta_{I,J}\frac{\tau}{\tau+|I|}\,.
\label{strongY}
\end{equation}
To make a prediction about the presence or absence of cluster formation, we compute the distribution of Hamming distance for this case. Inserting the result (\ref{strongY}) into equation (\ref{XiY}) we obtain
\begin{eqnarray}
&\big\langle\Xi(n)\big\rangle_\infty=\frac{1}{2^N}{N \choose n}\sum_{I,J}\Pi_{I,\{1,\ldots,n\}}\Pi_{J,\emptyset}\,Y_{IJ}\nonumber\\
&=\frac{1}{2^N}\sum_{m=0}^n\sum_{k=0}^{N-n}{N \choose n}{n\choose m}{N-n\choose k}(-1)^m\frac{\tau}{\tau+k+m}\nonumber\\
\noalign{\medskip}&=\frac{\Gamma(N+1)\,\Gamma(\tau+1)\,{_2F_1}(n-N,\tau;n+\tau+1;-1)}{2^N\,\Gamma(N-n+1)\,\Gamma(n+\tau+1)}\,,
\label{2F1}
\end{eqnarray}
where ${_2F_1}$ denotes the hypergeometric function. The last line is established by using the integral representation $\alpha^{-1} = \int^{\infty}_{0} e^{-\alpha z}\,dz$ for $\alpha = \tau + k+m$, after which the two sums become simple binomial expansions.\par
Depending on the value of $\tau$, equation (\ref{2F1}) predicts that the distribution of pairwise Hamming distance will interpolate between a symmetric binomial and a delta function at zero. This is illustrated in Fig.~\ref{strongfig}, which shows the distributions resulting from the values $\tau=1$ and $\tau=4$. In both cases the deterministic theory predicts a binomial distribution, as the values $\mu$ are well within the stable region (see set of points `3' in Fig.~\ref{phased}). As $\tau$ increases, the left skew of the distribution becomes stronger, meaning that the population has grouped together into tight clusters of genotypically similar organisms. Clusters have formed.  \par
Whilst the agreement between simulations and theory for the \textit{average} distribution of pairwise Hamming distance is excellent, we should point out that the measured distributions vary greatly from one simulation run to the next. We demonstrate this in the figure by plotting the results of the first simulation in both samples. The presence of multiple peaks implies the formation of several disjoint clusters. 

\section{Conclusion}

To summarise, we have investigated a simple individual-based genetic evolutionary model, which is driven by the effects of mutation and competition. Theoretical analysis in the limit of large population size revealed several interesting phenomena. On a macroscopic (deterministic) level, the model exhibits a pattern-forming transition whereby the decline of competition strength with genetic distance can drive the formation of genotypic clusters in an initially diverse population. On further investigation it was found that this pattern-forming process is highly amplified by the effects of demographic noise in the model. Large but finite populations exhibit quasi-clustering when the mutation strength is relatively large while, more strikingly, lower mutation strengths lead to the formation of clearly distinct clusters which are not predicted by the deterministic analysis.  We have demonstrated that the propensity to form clusters is determined by the average total number of mutations per generation in the 
population. 

The phenomenon of spontaneous speciation was first observed in the phenotypic version of the model \cite{Rogers2012}. In fact, whilst the combinatorial aspects are more involved, the essential flavour of the calculation presented here is the same. What we have achieved by introducing a genetic formulation is a step towards greater biological relevance, as well as providing further evidence that this mechanism of speciation is both general and robust. In forthcoming work \cite{Rossberg2012}, we will examine the implications of this thesis in the wider context of population genetics.

The biological relevance of the work could be further improved by consideration of a number of features which have been omitted from the model. These include: epistasis, and more generally the complex relationship between genotype and phenotype; sexual reproduction and the emergence of reproductive isolation of species; heterogeneity in the fitness landscape; geographic distribution of the population leading to allopatric/parapatric speciation. Inclusion of any of these features would provide a useful generalisation of the model. It is worth pointing out, however, that such considerations will not overturn our basic finding that demographic noise is itself a fundamental force in the process of speciation. 

\section*{Acknowledgements}
TR acknowledges funding from the EPSRC under grant number EP/H02171X/1.  
\section*{Bibliography}
\bibliographystyle{iopart-num}
\bibliography{SGCPCO_refs}

\appendix
\section{Properties of $\Pi$}
In the main text we claimed that the matrix $\Pi$ with entries $\Pi_{IJ}=(-1)^{|I\cap J|}$ diagonalises any matrix whose entries are functions of Hamming distance. Before proving this, we demonstrate some other useful properties of $\Pi$. Firstly, for any $I,J$ and $K$ we have the identity
\begin{equation}
\Pi_{IK}\Pi_{KJ}=\Pi_{K,I\ominus J}\,.
\label{niceGamma}
\end{equation}
To see this, one must simply observe that 
\begin{equation*}
|K\cap I|+|K\cap J|=2\,|K\cap (I\cap J)|+|K\cap (I\ominus J)|\,.
\end{equation*}
Secondly, 
\begin{equation}
\sum_{K} \Pi_{IK}\,\Pi_{KJ} = 2^{N}\,\delta_{IJ},
\label{invPi}
\end{equation}
which implies that $\Pi$ is a multiple of its own inverse: $\Pi^{-1}=2^{-N}\,\Pi$. This follows from (\ref{niceGamma}), and the fact that 
\begin{equation*}
\sum_{K}\Pi_{KL}=2^N\,\delta_{L,\emptyset}\,.
\end{equation*}
This last relation can be seen to be true by noting that the rows of $\Pi$ are sequences of plus ones and minus ones, with equal numbers of each --- except for the first row, which is all ones.

Now, suppose $F$ is a matrix whose entries are determined by Hamming distance according to $F_{IJ}=f(|I\ominus J|)$ for some function $f$. We compute
\begin{eqnarray}
\frac{1}{2^N}\Big[\Pi F\Pi\Big]_{IJ}&=\frac{1}{2^N}\sum_{K,L}\Pi_{IK}f(|K\ominus L|)\Pi_{LJ}\nonumber\\
&=\frac{1}{4^N}\sum_{K,L,M}\Pi_{IK}\Pi_{K\ominus L,M}\widetilde{f}(|M|)\Pi_{L,J}\nonumber\\
&=\frac{1}{4^N}\sum_{K,L,M}\Pi_{IK}\Pi_{KM}\Pi_{LM}\Pi_{L,J}\widetilde{f}(|M|)\nonumber\\
&=\delta_{I,J}\widetilde{f}(|I|)\nonumber\,.
\end{eqnarray}
Here the second line follows from application of the inverse transform defined in (\ref{invf}); the third by the property (\ref{niceGamma}); and the fourth from the sums over $K$ and $L$ collapsing to $2^N\delta_{I,M}$ and $2^N\delta_{J,M}$, respectively, according to (\ref{invPi}).

As a useful example, we compute the transform of the mutation matrix $R_{IJ}$ defined in 
Eq.~(\ref{defR}). Writing $R_{IJ}=r(|I\ominus J|)$, where $r(n)=\mu^n(1-\mu)^{N-n}$, the above calculation provides
\begin{eqnarray}
\frac{1}{2^N}\Big[\Pi R\Pi\Big]_{IJ}=\delta_{IJ}\,\widetilde{r}(|I|)\,.
\end{eqnarray}
The transformation of $r$ may be performed explicitly:
\begin{eqnarray}
\widetilde{r}(|I|) &=& \sum_{J} 
\Pi_{IJ} \mu^{|J|}\left( 1 - \mu \right)^{N-|J|} \nonumber \\
&=& \sum^{|I|}_{k=0}\,\sum^{N-|I|}_{\ell = 0} {|I|\choose k} {N-|I|\choose \ell} 
\left( -1 \right)^{k} \mu^{k+\ell}\left( 1 - \mu \right)^{N-k-\ell} \nonumber \\
&=& \sum^{|I|}_{k=0} {|I|\choose k} \left( -1 \right)^{k} \mu^{k}
\left( 1 - \mu \right)^{|I|-k} \nonumber \\
&=& \left( 1 - 2\mu \right)^{|I|}.
\end{eqnarray}
The second line was obtained from the first by decomposing the sum over the 
sets $J$ into the process of choosing $k$ elements from $I$ and $\ell$ from 
the compliment, to form a set of size $k+\ell$.

\section{Conditioned Stochastic Differential Equations}
In our calculation for the strong-noise regime, we reduced the number of stochastic degrees of freedom in the system by enforcing the condition $y_\emptyset\equiv 1$. In this Appendix we show how conditioning a stochastic differential equation (SDE) in this way alters the covariance structure of the noise experienced by the other variables. Applied to our system, the general derivation given here leads to Eq.~(\ref{condy}) in the main text.\par
Consider a vector of variables $\bm{x}=(x_0,x_1,\, \ldots\,, x_n)$, satisfying the SDE
\begin{equation}
\frac{d\bm{x}}{dt}=\bm{F}(\bm{x})+G(\bm{x})\bm{\eta}(t)\,,
\label{SDE}
\end{equation}
where $\bm{\eta}(t)$ is a vector of independent Gaussian white noise variables, and $\bm{F}$ and $G$ are vector- and matrix-valued functions of the state $\bm{x}$, respectively. Alternatively, we could have written the equivalent formulation
\begin{equation}
\frac{d\bm{x}}{dt}=\bm{F}(\bm{x})+\bm{\zeta}(t)\,,
\end{equation}
where $\bm{\zeta}(t)$ is a vector of \emph{correlated} Gaussian white noise variables, with covariance matrix $B=GG^T$. That is, 
\begin{equation*}
\langle\zeta_i(t)\zeta_j(t')\rangle=\delta(t-t')B_{ij}\,.
\end{equation*}

Suppose we wish to impose upon the system the condition $x_0\equiv c$, for some constant $c$. We write $\bm{x}_*=(x_1,\, \ldots\,, x_n)$ for the remaining degrees of freedom, and aim to derive an SDE for their behaviour under the constraint. 

First, applying the Gram-Schmidt process to $G(\bm{x})$ we can always write $G(\bm{x})=LQ$, where $L$ is lower-triangular, $Q$ is orthogonal, and both depend on $\bm{x}$ (although we have suppressed this in the notation). We separate $L$ into the parts relevant to $x_0$ and $\bm{x}_*$ by writing it in block form
\begin{equation*}
L=\left(\begin{array}{cc}L_{00}&0\\L_{*0}&L_{**}\end{array}\right)\,,
\end{equation*}
where $L_{00}$ is $1\times 1$, $L_{*0}$ is $n\times 1$ and $L_{**}$ is $n\times n$. Note that $B=GG^T=LL^T$, so writing $B$ in block form also we obtain  
\begin{equation}
\left(\begin{array}{cc}B_{00}&B_{0*}\\B_{*0} &B_{**}\end{array}\right)=\left(\begin{array}{cc}L_{00}^2&L_{00}L_{*0}^T\\L_{00}L_{*0} &L_{*0}L_{*0}^T+L_{**}L_{**}^T\end{array}\right)\,.
\end{equation}
 
Applying the transformation $Q$ to the vector of noise variables, we write $\bm{\sigma}(t)=Q\bm{\eta}(t)$. It is known that for any such state-dependent orthogonal transformation of Gaussian white noise, the transformed process $\bm{\sigma}(t)$ has the same statistics as the original $\bm{\eta}(t)$ (see, for example, \cite{Gardiner1985}). In our case we deduce that
\begin{equation*}
\frac{d\bm{x}}{dt}=\bm{F}(\bm{x})+L\bm{\sigma}(t)\,,
\end{equation*}
where $\langle \sigma_i(t)\sigma_j(t')\rangle=\delta_{i,j}\delta(t-t')$. Finally, imposing $x_0=c$, we obtain
\begin{equation}
\sigma_0(t)\equiv-\frac{F_0(\bm{x})}{L_{00}}\Big|_{x_0=c}\,.
\label{B4}
\end{equation}
For the remaining degrees of freedom, we arrive at 
\begin{equation}
\frac{d\bm{x}_*}{dt}=F_*(\bm{x})\Big|_{x_0=c}-L_{*0}\frac{F_0(\bm{x})}{L_{00}}\Big|_{x_0=c} + L_{**}\bm{\sigma}_*(t)\,,\Bigg.
\end{equation} 
or equivalently 
\begin{equation}
\frac{d\bm{x}_*}{dt}=F_*(\bm{x})\Big|_{x_0=c}-F_0(\bm{x})\frac{L_{*0}}{L_{00}}\Big|_{x_0=c} + \bm{\zeta}_*(t)\,,
\end{equation}
where the correlation matrix for the noise variables $\bm{\zeta}_*$ is simply $L_{**}L_{**}^T$, or, in terms of the original correlation matrix $B$:
\begin{equation}
L_{**}L_{**}^T=B_{**}-\frac{B_{*0}B_{0*}}{B_{00}}\,.
\end{equation}
To obtain equations (\ref{langevinS}) and (\ref{condy}) in the main text, we apply the condition $y_\emptyset\equiv1$ to the system (\ref{oldsystem}), with the $B$ matrix specified by Eq.~(\ref{oldnoise}). Note that in this case the right-hand side of Eq.~(\ref{B4}) is zero.

\end{document}